\documentclass{emulateapj}
\usepackage{color}

\slugcomment{Draft,\today}

\def\be{\begin{equation}}
\def\ee{\end{equation}}
\def\kms{{\rm\,km s^{-1}}}
\def\bh{_{\bullet}}
\def\bulge{_{\rm *,bulge}}
\def\min{{\rm min}}
\def\msun{M_{\odot}}
\def\lsun{L_{\odot}}
\def\mrg{_{\rm mrg}}
\def\tot{_{\rm tot}}

\def\kpc{{\rm\,kpc}}
\def\Gyr{{\rm\,Gyr}}
\def\yr{{\rm\,yr}}

\def\ergs{{\rm\,erg~s^{-1}}}
\def\sun{_{\odot}}

\begin{document}
\title{
The low frequency of dual AGNs versus the high merger rate of galaxies:
A phenomenological model
}
\author{Qingjuan Yu$^{1}$, Youjun Lu$^{2}$, Roya Mohayaee$^{3}$ 
and Jacques Colin$^{3}$}
\affil{~$^{1}$Kavli Institute of Astronomy and Astrophysics, Peking 
University, Beijing 100871, China; yuqj@pku.edu.cn\\
~$^{2}$National Astronomical Observatory of China, Beijing 100012, China\\
~$^{3}$UPMC, CNRS, Institut d'Astrophysique de Paris, 98 bis Bd. Arago, 
Paris 75014, France
}

\begin{abstract}

Dual AGNs are natural byproducts of hierarchical mergers of galaxies
in the $\Lambda$CDM cosmogony. Recent 
observations have shown that only a small fraction 
($\sim 0.1\%-1\%$) of AGNs at redshift
$z\la 0.3$ are dual with kpc-scale separations, which 
is rather low compared to the high merger rate of galaxies. Here we construct 
a phenomenological model to estimate
the number density of dual AGNs and its evolution according to the 
observationally-estimated major merger rates of galaxies and 
various scaling relations on the properties of galaxies and their central massive black holes.
We show that our model reproduces the observed frequency and 
separation distribution of dual AGNs provided that 
significant nuclear activities are triggered only in gas-rich progenitor
galaxies with central massive black holes and only when the nuclei of these galaxies are 
roughly within the half-light radii of their companion galaxies.
Under these constraints, the observed low dual AGN
frequency is consistent with the relatively high merger rate of galaxies 
and supports the hypothesis that major
mergers lead to AGN/QSO activities. We also predict that the number of kpc-scale dual AGNs
decreases with increasing redshift and only about 
0.02\%--0.06\% of AGNs are dual AGNs with double-peaked narrow line features at redshifts of $z\sim
0.5-1.2$. Future observations of high-redshift dual AGNs would 
provide a solid test for this prediction.

\end{abstract}


\section{Introduction }\label{sec:intro}

Merger of galaxies is one of the major processes of galaxy formation in the
hierarchical $\Lambda$CDM cosmogony. Galaxy mergers can naturally lead to the
formation of paired and binary massive black holes (MBHs), since most galaxies,
especially those with spheroidal components, host MBHs at their centers
\citep[e.g.,][]{BBR80,Yu02}. If both of the merging galaxies are gas rich, a
large amount of gas can be channeled to the central region of each merging
galaxy, as indicated by numerical simulations \citep[e.g.,][]{Hernquist89}.
A dual AGN (dAGN) could then emerge
if the accretion onto both MBHs is triggered during the merging process.  
Thus the existence of dAGNs in the
universe and their demography provide an important probe not only of the
hierarchical galaxy formation models but also of the 
triggering mechanisms of nuclear activities and the assembly history of MBHs. 

In the past decade, substantial progress has been made 
in searching for dAGNs and binary MBHs
(BBHs) through various possible signatures, such as
double-peaked broad lines \citep[e.g.,][]{BL09}, double-peaked narrow
lines \citep[e.g.,][]{Zhouetal04,Gerke07,Comerford09, Comerford09b,Xu09,
Wangetal09,Smith10, Liu10a,Liu10b,Shen10,Fu10,Rosario11} 
and various other methods
\citep[e.g.,][]{Komossa03, Valtonen08, Green10}. 
Following a systematic
scrutiny of the NIR images and optical slit spectra of a sample of
double-peaked narrow line type 2 AGNs, 
it has recently been reported that 
roughly 0.5\%-2.5\% of the $z\la 0.3$ 
type 2 AGNs are kpc-scale dAGNs after taking into account 
the selection completeness \citep[][see also \citealt{Rosario11}]{Liu10a,Liu10b,Shen10}. 
This frequency is surprisingly low even if we were to assume that
all these observations are indeed of dAGNs and not as 
has been otherwise suggested of superposed single AGNs, bipolar jets and accretion disks.
The observed frequency of dAGN is indeed more than an order of magnitude lower than that 
expected from the observed major merger
rate of galaxies if one makes the assumption 
that each major merger would yield a dAGN \citep[][see also
\citealt{Shen10}]{Rosario11}. The low frequency of dAGNs, unlikely due to
selection effect \citep{Rosario11}, 
seems generally at odds with the commonly
accepted scenario that AGN/QSO activities are triggered by major mergers of
galaxies.

In this paper, we construct a phenomenological model to address the observed
frequency of kpc-scale dAGNs, by taking into account the following factors: (1) the merger
rate of galaxies; (2) the types of galaxy mergers that can trigger the nuclear
activities of both progenitors; (3) when and where significant nuclear activities are
triggered (by ``significant'' nuclear activities we mean that the Eddington ratios of the 
nuclear luminosities are close to 1 (e.g., 0.1--1), rather than $10^{-3}$ or less);
and (4) how long the nuclear activities of both progenitors can last
before the merger of the two MBHs. After the merger of two MBHs, the nucleus
of the merged galaxy is likely to still be active, but it only appears as a
single AGN.
In Section~\ref{sec:model}, we show how these factors are adopted in our model
of the dAGN distribution. Under the assumption that significant nuclear activities can be 
triggered only in gas-rich progenitor galaxies with central MBHs when their
companion galaxies are sufficiently close to them, the dAGN frequency and their separation 
distribution are obtained in Section~\ref{sec:results}. 
By comparison with observations, we show that the observational frequency of dAGNs
is consistent with current observational constraints on the merger rates of
galaxies and the scenario that major mergers of galaxies lead to significant nuclear
activities. 
In Section~\ref{sec:results}, we 
also present our model predictions
for the dAGN distribution at higher redshifts. We further discuss the
uncertainties in our estimates of the dAGN frequency due to
different processes, which are not included in our model, such as mergers of progenitor galaxies whose nuclei
have already been activated in previous mergers and also
tidally-induced nuclear activities in two galaxies which 
are still far away from each other.
Conclusions are given in Section~\ref{sec:conclusions}.

\section{A phenomenological model of dual AGNs }\label{sec:model}

In this section, we introduce a phenomenological model to estimate the
frequency of dAGNs, by taking into account the factors mentioned in
Section~\ref{sec:intro}. The crucial elements of this model are described as
follows. 

\subsection{Merger rates of galaxies}\label{subsec:mrg}

For galaxies with stellar mass $\geq M_*$, we describe their merger rate at a
given time by the number fraction of those galaxies for which a
galaxy merger completed per unit time, and denote the merger rate by
\be
{\cal R}(\geq M_*,x,z)\equiv \frac{1}{n\tot(\geq M_*,z)}
\frac{dn\mrg(\geq M_*,x,z)}{dt_{z}},
\label{eq:mrg}
\ee
where $t_{z}= \int^{\infty}_{z} |\frac{dt}{dz'}| dz'$ is the cosmic time at
redshift $z$, $x$($\le 1$) represents the mass ratio of two merging galaxies,
$n\tot(\geq M_*,z)$ is the comoving number density of
galaxies with stellar mass $\geq M_*$ at redshift $z$, and $(dn\mrg/dt_z)dt_z$ gives the comoving number density of those galaxies
that are products of mergers of two progenitor galaxies with mass ratio $\ge x$
and with their mergers being completed over a cosmic time from $t_z$ to $t_z+dt_z$. 

In the past several years, tremendous efforts have been made in estimating the
merger rate of galaxies either through close pairs of galaxies
\citep[e.g.,][]{Kartaltepe07, Lin04, Lin08, dePropris05, Bell06a, Bell06b,
Bundy09, LS09, dR09, Patton11} or through morphological disturbances of galaxies
found in various deep surveys \citep[e.g.,][]{Cassata05, Bundy05, Wolf05,
Bridge07, Bridge10, Lotz08, Conselice08, Conselice09}. The dependence of
the merger rate ${\cal R}(\geq M_*,x,z)$ on mass ratio 
$x$ can be absorbed in a function $f(x)$, which describes 
the fraction of mergers with mass ratio larger than $x$, i.e.,
\be
{\cal R}(\geq M_*,x,z) =f(x) R(\geq M_*,z).  
\label{eq:R}
\ee
For major mergers (usually defined by $x\geq 1/3$), observational
estimates can be roughly fitted by the simple formula, 
$
R(\geq M_*,z)=A(M_*)_{\rm major}(1+z)^{\beta(M_*)_{\rm major}}, 
$
where
$
A(M_*)_{\rm major} \approx 0.2[1+(M_*/M_0)^{0.5}]\Gyr^{-1}, 
$
$
\beta(M_*)_{\rm major} \approx1.65-0.15\log(M_*/M_0), 
$
and $M_0=2\times 10^{10}\msun$ \citep{Hopkins10}.
The normalization of the merger rate $A(M_*)$ is 
uncertain by a factor of about $2$ due to systematic errors and
the uncertainty in the evolution slope $\beta(M_*)$ is $\Delta\beta \sim
0.15-0.20$. The dependence of the merger rates on mass ratio $x$ can be
approximated by $f(x)\propto x^{-0.3}(1-x)$ \citep{Stewart09,Hopkins10}.

The stellar mass function of galaxies, involved in equation (\ref{eq:mrg}), has
been estimated over a large redshift range from various galaxy redshift
surveys. In our Monte-Carlo calculations below, we adopt the latest stellar
mass function obtained from the S-COSMOS survey (Spitzer-Cosmic Evolution
Survey) for
galaxies at redshift ranges $0.2 - 0.4$, $0.4-0.6$, $0.6-0.8$, $0.8-1.0$, and
$1.0-1.2$ \citep{Ilbert10}, and that obtained from the Sloan Digital Sky Survey
(SDSS) for galaxies in the local universe ($z\sim 0.1$) \citep{Bernardi10},
respectively. According to equation (\ref{eq:mrg}) and
the stellar mass functions, our simulations generate a large number of merging
pairs of galaxies over redshift $0$ to $1.2$. The masses of the two progenitors are
assigned by $M_{*,1}=\max(\frac{xM_*}{1+x},\frac{M_*}{1+x})$ and 
$M_{*,2}={\rm min}(\frac{xM_*}{1+x},\frac{M_*}{1+x})$, respectively, where $x$ is randomly
selected according to its distribution function $f(x)$.\footnote{
Note that a substantial
number of stars may be formed in the merging systems during the merging
process. We have checked that if a fraction (e.g., $10\%-50\%$) of
the total stellar mass $M_*$ is removed in order to get the real total
stellar mass of the two progenitor galaxies before doing the splitting procedure
above, a slightly lower estimate of the dAGN frequency may be obtained, but
which does not affect our conclusions in Section~\ref{sec:conclusions}. }

\subsection{Triggering of nuclear activity and mergers of galaxy pairs with
various morphologies}

During the merging process of two galaxies, whether significant nuclear
activities at their centers can be triggered depend on the two factors: (1)
whether MBHs are initially hosted in both nuclei or whether the MBHs are
massive enough; and (2) whether sufficient gaseous materials can be quickly
delivered into the vicinity of the MBHs. Those two factors are closely related
to the progenitor morphologies of the merging pairs.

\subsubsection{The mass of the initial MBHs}
\label{subsec:mbh}

Observations have shown that the masses of MBHs, $M\bh$, in the centers of nearby
galaxies are tightly
correlated with the stellar mass of the spheroidal components of the galaxies
$M\bulge$ \citep{HR04}
\be
\log M\bh=8.28+1.12(\log M\bulge-11),
\label{eq:bhbulge}
\ee
with an intrinsic scatter of $0.3$~dex in $\log M\bh$. Subsequently, given the stellar
mass of the spheroidal component $M_{*,{\rm bulge},i}$ $(i=1,2)$ of each progenitor of a
merging pair, we use equation (\ref{eq:bhbulge}) to estimate the central BH
mass $M\bh{_{,i}}$ ($i=1,2$) in the progenitor at redshift $z$, by assuming that the
scatter of the correlation follows a normal distribution and adopting a small
evolution correction of $M\bh \propto (1+z)^{0.68}$ \citep[e.g.,][]{Merloni10}.

Note that the following elements are involved to estimate the distribution
of $M\bulge$:

(a) The bulge to total stellar mass ratio (B/T):
given the stellar mass $M_*$ of a galaxy, the mass $M\bulge$ can be estimated
according to the morphology of the galaxy, as the B/T ratios
are different for galaxies with different
morphologies \citep{Weinziri09}. Based on the detailed 
analysis of a sample of nearby galaxies,
it has been found that $B/T=0.22$ and $0.05$ with variances of $0.05$ and
$0.02$ for Sa-Sb and Sc-Sd, respectively \citep{Weinziri09}. We 
assume that $B/T=1$ for elliptical
and S0 galaxies and $B/T=0$ for irregular galaxies, respectively. 

(b) Distribution of morphological combinations of merging galaxy pairs:
it is possible that the two progenitors of a merging galaxy pair have
different morphologies. Currently detailed statistics on the morphological
combinations of the pairs are not available, although there are quite a
number of constraints on the merger rates of red galaxies, blue galaxies and
mixed ones \citep[e.g.,][]{Lin04, Lin08}.  Here we simply assume that the
morphological types of the two progenitors are independent of each other, and
consequently the fraction of the merger rates of the pairs with any specific
morphological combination to the total merger rates only depends on the
fraction of these types to the total number of galaxies.  This assumption
appears consistent with the current observations that the fractions of
different morphological types for 
galaxies in pairs are similar to those for
field galaxies \citep{dePropris05}. 

(c) Stellar mass functions of galaxies with different morphological types:
the galaxy luminosity function (LF) for four spectral
types of galaxies over $z\sim 0.2-1.2$ was previously estimated
(see Table 3 of \citealt{Zucca06}). These four spectral
types roughly correspond to the morphological types E/S0, Sa-Sb, Sc-Sd
and Irr, respectively. For each type of galaxies, the mass-to-light ratio can
be estimated through their average colors \citep[e.g., B-V, see][]{Bell03,
Bernardi10, Fukugita95}, and their LFs \citep{Zucca06} can thus be converted to
the stellar mass functions.  According to these mass functions, the fraction of
each type of galaxies to the total galaxies can be obtained at any given $M_*$
over $z\sim 0.2-1.2$. For galaxies at redshifts $z<0.2$, we adopt the stellar
mass functions for different morphological types given by \citet[][see their
Table B2]{Bernardi10} and estimate the fraction of each type to the total at
any given $M_*$.  In our calculations below, for each merger generated as that
described in Section~\ref{subsec:mrg}, we randomly set the morphological type
to each progenitor according to the fractions of each type of galaxies to the
total at $M_{*,1}$ and $M_{*,2}$, respectively.

For those irregular or purely disk-like galaxies, we set the initial
mass of their MBHs to be $10^5\msun$ \citep{Ho08,Kormendy11} or zero.
Here, these two different initial values do not 
lead to significantly different results.  

\subsubsection{Triggering of the nuclear activity}\label{subsec:dc}

During the merger of late-type gas-rich galaxies (Sa-Sb,
Sc-Sd, or Irr) with other galaxies, significant nuclear activities 
may be triggered rapidly. The reason lies in the fact that late-type
galaxies contain significant amount of gaseous materials 
which could lose angular momentum and sink
under dynamical friction during the
galaxy merger. Early-type gas-poor (red/elliptical) galaxies may be able to
capture some gas from gas-rich encounters during the merging process, but the
time for the gas to reach the center is likely to be comparable to the merging
timescale (see Section~\ref{sub:Devol}). Therefore, the nuclear activities in
these galaxies may start only after the mergers have almost completed. In addition,
the host galaxies of all the confirmed dAGNs selected through double-peaked
narrow lines contain stellar disks \citep{Shen10}, which would suggest that
they are merging remnants of late-type galaxies.

Based on the above arguments, we assume that the nuclear activity is triggered
in each gas-rich component with a central MBH once its separation with its
companion becomes smaller than a threshold $D_{\rm c}$. On the other hand, we assume 
that the nuclear activity cannot be triggered in any gas-poor component of 
a merging pair prior to the completion
of the merger (see also discussions in Section~\ref{sec:results}). The physical size of a galaxy is characterized by its half-light
(or effective) radius $r_{\rm h}$. Once the separation of the component $i$
(1 or 2) of a pair to the other component $j$ (2 or 1) is smaller than
the half-light radius of component $j$ (i.e., $r_{{\rm h},j}$), the center of
component $i$ may be significantly perturbed. Here we assume the
threshold $D_{\rm c}$ for component $i$ to be $D_{{\rm c},i}= Kr_{{\rm h},j}$,
where $K$ is a fudge factor of order unity.  For major mergers,
nuclear activities can be triggered in both components only if both components
are gas-rich and their separation $D\leq D_{\rm c}= K\min(r_{\rm h,1},r_{\rm
h,2})$, where $r_{\rm h,1}$ and $r_{\rm h,2}$ are the half-light radii of the
primary and the secondary galaxy of the pair, respectively.  

For disk galaxies, we adopt the following empirical relation to estimate the
disk half-light radius from the stellar mass, which is obtained from a SDSS
sample at low redshift $z\sim 0.1$ \citep{Dutton10a}
\be
r_{\rm h}=r_0\left(\frac{M_*}{M_0}\right)^{\alpha}\left[\frac{1}{2}+\frac{1}{2}
\left(\frac{M_*}{M_0}\right)^{\gamma}\right]^{(\beta-\alpha)/\gamma},
\label{eq:rdisk}
\ee
where $\alpha=0.18$, $\beta=0.52$, $\gamma=1.8$, $r_0=5.2\kpc$ and $M_0=2.75
\times 10^{10}\msun$. The scatter of this relation at a given $M_*$ is assumed
to follow the log-normal distribution with a standard deviation of
$\sigma=s_2+(s_1-s_2)/[1+(M_*/M^{'}_0)^{\eta}]$,
where $s_1=0.47$, $s_2=0.27$, $M^{'}_0=2\times 10^{10}\msun$ and $\eta=2.2$.
For disk galaxies at higher redshifts, for example $z\sim 0.2-1.2$, we add an evolution
correction to the zero point of the disk-size---stellar mass relation by
$\Delta \log r_0=0.018-0.44 \log(1+z)$ \citep{Dutton10a}.

For early-type galaxies, the half-light size versus stellar mass relation can also be
fitted by equation (\ref{eq:rdisk}), but with $\alpha=0.03$, $\beta=0.64$,
$\gamma=1.3$, $r_0=1.4\kpc$, and $M_0=1.2\times 10^{10}\msun$
\citep{Dutton10b}. This double power-law form is consistent with the single
power-law form at high masses given by \citet{Wel08} and \citet{Shen03}. The
intrinsic scatter of this relation at high masses is $\sim 0.14$~dex
\citep{Wel08}. We assume the scatter is also $\sim 0.14$~dex at low masses
($\sim 10^9-10^{10}\msun$).  For early-type galaxies at higher redshifts, for example
$z\sim 0.2-1.2$, we add an evolution correction to the zero point of this
relation as $\Delta\log r_0=-0.98\log (1+z)$ \citep{Wel08}.

\subsection{Evolution of the separation of two merging galaxies}
\label{sub:Devol}

The merging timescale of paired galaxies is crucial in determining the time
period during which the merging system would appear as dAGNs. Based on virtual galaxy
catalogues obtained from the Millennium Simulation, it has been found that
the average merging time of paired galaxies can be fitted by
\begin{eqnarray}
\langle T\mrg\rangle (r_{\rm p}) & = & 2.2\Gyr \left(\frac{r_{\rm p}}{50\kpc}\right) 
\left(\frac{M_*}{4\times 
10^{10}h^{-1}\msun} \right)^{-0.3} \nonumber \\
& & \times \left(1+\frac{z}{8}\right),
\label{eq:tmrg}
\end{eqnarray}
where $r_{\rm p}$ is the projected 
separation of the pairs \citep{KW08}. This result is
about $15\%-30\%$ larger than that obtained from higher
resolution simulations of galaxy mergers \citet{Lotz10}. Correcting this difference and
assuming that the orientations of galaxy pairs are isotropically distributed,
the merging time as a function of three-dimensional separation $D$ is
smaller than that given by equation (\ref{eq:tmrg}) for a projected separation
$r_{\rm p}=D$ by about a factor of $2$, i.e., 
\be
\left <\tau\mrg\right>(D)\simeq
0.5\left<T\mrg\right>(r_{\rm p}=D).
\label{eq:tDmrg}
\ee

The period that a merging system may appear as a double-peaked narrow line dAGN
varies from $\delta t=0$ to $\delta t=\tau\mrg(D_{\rm c})-\tau\mrg(D_{\rm L})$, 
where $D_{\rm L}$ is the lower limit of the separation
that the narrow line regions associated with the two nuclei begin to overlap and
the line profile becomes more complicated, and $\delta t$ is the time elapsed
since the activities of both the nuclei were triggered.
Here we set $D_{\rm L}=0.5\kpc$
since the typical size of narrow line regions of AGNs is about a few tens to a few
hundreds of parsecs \citep{Peterson}.

\subsection{Luminosity evolution of nuclear activities}\label{subsec:lumevol}

Once the nuclear activity is triggered in the center of a merging component,
the material available for accretion initially is expected to be abundant. The
situation of these merging systems may be similar to that of high luminosity
QSOs, which also have plentiful gas supplies and accrete at relatively high rates.
The Eddington accretion rate is $\dot{M}_{\rm Edd}\simeq 0.22\msun\yr^{-1}
(\epsilon^{-1}-1)(M\bh/10^8\msun)$, where $\epsilon \sim 0.1-0.16$ is the
mass-to-light conversion efficiency \citep[e.g.,][]{YT02,YL08}. Here we assume
that the accretion rate is $\lambda \dot{M}_{\rm Edd}$ for the MBH in each
component of a merging pair after its nuclear activity has been triggered, and we
set $\lambda=0.25$ (e.g., as shown in \citealt{Kollmeier06,Shen08}) unless 
otherwise stated. Thus the MBH
should grow exponentially with the elapsed time as $\exp(\lambda \delta
t/\tau_{\rm Sal})$, where the Salpeter timescale $\tau_{\rm Sal}$ is about
$4.5\times 10^8(\frac{\epsilon}{1-\epsilon})\yr$.
We assume that the accretion switches off if the total mass of the
two MBHs is larger than the expectation from equation (\ref{eq:bhbulge}) for
the final merged galaxy, taking into account the scatter in the relation.\footnote{
Even if these BHs can still accrete some material via lower Eddington ratios, e.g., $\sim 0.001$,
our simulation results are not affected much.} 
The bolometric luminosity is
$L_{\rm bol}=\lambda L_{\rm Edd}(M\bh)$, where $L_{\rm Edd}\simeq
1.3\times 10^{38}\ergs (M\bh/\msun)$. For the purpose of this paper, we need
calculate the [OIII] luminosity, which can be converted from the bolometric
luminosity by $L_{\rm [OIII]} \sim L_{\rm bol}/3500$ with a scatter of $\sim
0.38$~dex (assumed to have a log-normal distribution) \citep{Heckman}. 

\subsection{Velocity separation}\label{subsec:vsep}

In recent systematic searches, double-peaked narrow lines
have been adopted as the indicator to search for dAGNs 
from parent SDSS AGN samples \citep{Shen10}.
In this search the dAGNs with velocity separation $v_{\rm sep}<150\kms$ or $v_{\rm
sep} <\max(\sigma_{\rm e,1},\sigma_{\rm e,2})$ cannot be detected. 
The velocity dispersions of the
primary and the secondary progenitor galaxies (or their bulges) of the pair,
$\sigma_{\rm e,1}$ and $\sigma_{\rm e,2}$ respectively, characterize the 
widths of the two narrow line components.
In order to extract the dAGN frequency from these searches, it is
necessary to estimate $v_{\rm sep}$, $\sigma_{\rm e,1}$ and $\sigma_{\rm e,2}$
of the two merging components.  

As both components of a pair move in their common dark matter halo with a
velocity of the order of the circular velocity $v_{\rm circ}$ of the halo
(i.e.,  the merged galaxy), we first assume that the relative velocity of the two
components $v_{\rm rel}\sim 2v_{\rm circ}$ and later discuss the consequences of
this assumption in Section~\ref{sec:results}. It is plausible to assume
that the orbital orientation of the two merging components and their relative
position on the orbital plane are isotropically distributed. Subsequently the
velocity separation of the two components of the double-peaked lines, i.e., the
relative velocity component projected onto the line of sight, as well as their
projected separation perpendicular to the line of sight, can be taken into
account for each pair through
Monte-Carlo simulations.  For early-type galaxies, $v_{\rm circ}$ can be
estimated from  $v_{\rm circ}\sim 1.54\sigma_{\rm e}$ \citep{Dutton10b}. 
The velocity dispersion $\sigma_{\rm e}$
of the merger result which is supposed to be an elliptical, can be estimated from stellar
mass $M_*$ by the following empirical relation
\begin{eqnarray}
\log\left(\frac{\sigma_{\rm e}}{\kms}\right) =  2.23+0.37\log
\left(\frac{M_*}{10^{10.9}\msun}\right) \nonumber \\
 -0.19\log\left[\frac{1}{2}+
\frac{1}{2}\left(\frac{M_*}{10^{10.9}\msun}\right)
\right], 
\label{eq:vsig}
\end{eqnarray}
for nearby SDSS galaxies which has an intrinsic scatter of roughly
$0.071$~dex \citep{Gallazzi}.  For galaxies at higher redshifts, we add an
evolution correction to the above relation by interpolating its measured
deviations to higher redshifts \citep[][see their
Table~3]{Dutton10a}.  For a late-type galaxy, the velocity dispersion of its
bulge (if any) can also be estimated from equation (\ref{eq:vsig}) after
replacing $M_*$ there by $M\bulge$, as bulges appear to follow the same Faber-Jackson
relation as faint ellipticals \citep{BGP07}.

\begin{figure}
\epsscale{1.0}
%\epsscale{0.9}
\plotone{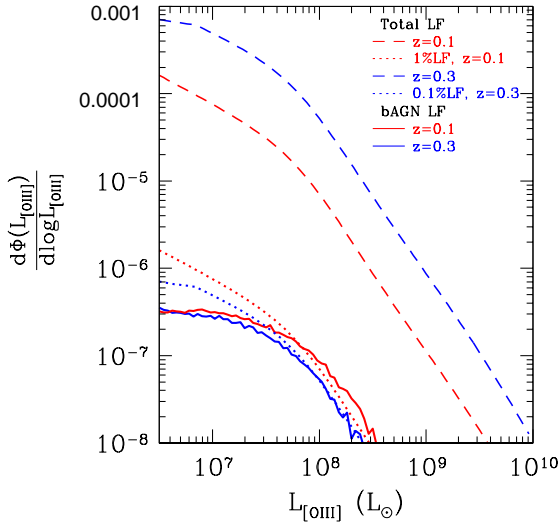}
%\centerline{\includegraphics[width=6.5in]{f1.eps}}
\caption{The LFs of AGNs and the bright components
of dAGNs. The red and blue dashed lines represent the observed [OIII] 
LF of all (including type 1 and type 2) AGNs at redshifts $z=0.1$ and $0.3$,
respectively \citep{Bongiorno10}. 
The dotted lines are reference lines.
The red dotted line represents 1\% of the observed AGN [OIII] LF
at $z=0.1$, and the blue dotted line represents 0.1\% of the observed AGN LF at $z=0.3$.
Each nucleus of a dAGN has its own luminosity, and thus the luminosity of a
dAGN is composed of two components. The red and blue solid lines represent
our model-predicted LF of the relatively bright components of dAGNs
at $z=0.1$ and $0.3$, respectively. Our simulated dAGNs are selected by
the criteria discussed in \S~\ref{subsec:vsep}. The typical error to the estimates of the dAGN LF
shown in this Figure is $0.5-0.6$~dex.
}
\label{fig:f1}
\end{figure}

\subsection{Monte-Carlo simulations of dAGNs}\label{ssubsec:MC}

We summarize the procedures to generate dAGNs in our Monte-Carlo simulations
as follows.

\begin{itemize}

\item We generate $10^7$ major mergers of galaxies which are completed during
the cosmic time $t_{z_k}$ to $t_{z_k}+\delta t_{z_k}$, where
$z_k=0.1, 0.3, 0.5, 0.7, 0.9$, and $1.1$ for $k=1, 2, ...,$ and $6$,
respectively, and $\delta t_{z_k}=10^9\yr$. The detailed completion time $t_z$
of a major
merger is selected according to the major merger rates and the stellar mass
function
given in Section~\ref{subsec:mrg}.  For each merger, the total stellar mass
$M_*$ ($10^9\msun-10^{12}\msun$) of the system is split into the stellar
masses of its two progenitor galaxies $M_{*,1}$ and $M_{*,2}$.

\item For each merging system, we assign morphological types to the two
progenitor galaxies and then estimate their bulge masses and the masses of
their central MBHs, as described in Section~\ref{subsec:mbh}.

\item For each progenitor galaxy of the system, the half-light radius is
estimated as described in Section~\ref{subsec:dc}, and their separation
threshold for triggering nuclear activities is correspondingly set to each
component, i.e., $D_{{\rm c},i}=K r_{{\rm h},j}$ ($i,j=1,2$ and $i\ne j$), where $K$ is
assumed to be the same for the two components of the major merger
and is calibrated by the separation distribution of observed
dAGNs (see Section~\ref{sec:results}).

\item We start from the completion time of each major merger and trace back the
evolution of the separation of its progenitor galaxies from zero to the values
at earlier time, as described in Section~\ref{sub:Devol}. (Note that our analyses
for kpc-scale dAGNs here are not affected by our extrapolation
of equation (\ref{eq:tDmrg}) to  pc scales and by our simplification of detailed
evolution timescales of massive bound binary black holes on these small scales.)
Assuming that all merging systems which complete their mergers within
the cosmic time $t_{z_k}$ to $t_{z_k}+\delta t_{z_k}$ are located at the
distance of redshift $z_k$, then these systems should appear in observations
at a time of $\Delta t=t_z-t_{z_k}$ before the completion of the
mergers, where $t_z$ is the cosmic time at which each major merger is
completed. For any merger, if
$\Delta t<\left<\tau_{\rm mrg}\right>(D_{{\rm c},i})$, the nuclear activity in 
component $i$ has been triggered, and the time elapse since the triggering of
the activity is given by $\left<\tau\mrg\right>(D_{{\rm c},i})-\Delta t$.

\item The luminosity is assigned to each active component as described
in Section~\ref{subsec:lumevol}.

\item If the nuclear activities in both components are triggered and the system
satisfies the velocity threshold set in Section~\ref{subsec:vsep}, it then
appears as a dAGN similar to those observed by \citet{Shen10}. 

\end{itemize}

In order to compare with the results given by systematic surveys by using
double-peaked narrow lines \citep[e.g.,][]{Shen10,Rosario11}, we only count the
number of those simulated dAGNs with two components having a separation within
the range from $0.5\kpc$ to $10\kpc$ and 
comparable luminosities,
i.e., the luminosity ratio of the two components is within a factor of 4 here.
And we then calculate their luminosity function and projected separation
distribution, etc., as illustrated in the following section.

\section{Model results and discussions}\label{sec:results}

Figure~\ref{fig:f1} shows the [OIII] LF of the bright components of the
simulated dAGNs that could be selected through double-peaked narrow lines (solid and long-dashed lines) at redshifts $z=0.1$ and $0.3$,
respectively.  These simulated dAGNs are selected through similar thresholds as
that in \citet{Shen10} for double-peaked narrow line dAGNs (see
Section~\ref{subsec:vsep}). And the threshold on the separation for the nuclear
activity to be triggered, $D_{{\rm c},i}= Kr_{{\rm h},j}$, is set to have
$K=1.25$ (which is the reference value, see also Figure~\ref{fig:f2}), where $i=1,2$
represent the two components of a merging pair and $i\ne j$. Note that the LF of the
faint components of dAGNs is only slightly smaller than that for the bright
components. The observed [OIII] LFs for all AGNs, including the [OIII] LFs of
the type 1 AGNs and type 2 AGNs adopted from \citet{Bongiorno10}, are also
shown in Figure~\ref{fig:f1}.  Apparently, the LF of the bright components of
dAGNs is smaller than that for all AGNs by a factor of about $100$ to $1000$.
Note that the bright component of an observed dAGN may be hosted in the
small progenitor galaxy \citep{Shen10}; and our model shows that about $30\%-40\%$ of
simulated dAGNs have their bright components being hosted in the progenitors with
relatively low total stellar masses, consistent with current observations.
Here we do not try to simultaneously fit the total AGN LF, as both
the detailed luminosity evolution of nuclear activities in the post-merger stages
and other mechanisms leading to nuclear activities would be important in such a fit
but beyond the scope of this paper.

The accuracy of our estimates of the dAGN LFs depends on the uncertainties
in (1) the merger rate ${\cal R}$, (2) the merger timescale $\left<\tau_{\rm
mrg}\right>(D)$, (3) the setting of the separation threshold $D_{\rm c}$, (4)
the setting of the accretion rate $\lambda$, and (5) the approximation of the
relative velocity. We discuss each of the points below.
\begin{itemize}

\item The systematic error of the merger rates, estimated from observations,
could affect our estimate by a factor of $2$. This error 
partly reflects the uncertainty in the
merger timescale \citep{Hopkins10}. The minor mergers of galaxies, not included
in our calculation, might also contribute to dAGNs. However, the rate of minor
mergers with mass ratio $x\in (1/10,1/3)$ is almost the same as the rate of major mergers
\citep{Hopkins10}. The contribution by these minor mergers should not be larger
than a few tenth of the contribution by the major merger of galaxies (with
$x\in (1/3,1)$),
as the two merger 
components differ substantially in stellar mass
and are less likely to contribute to AGN pairs 
with two comparably bright components.  Therefore,
the total error introduced by the uncertainties in ${\cal R}$ and
$\left<\tau_{\rm mrg}\right>(D)$ is roughly a factor of $2$.

\item The uncertainty in the settings of the separation threshold $D_{\rm c}$
may be calibrated by the observational separation distribution of dAGNs.
Figure~\ref{fig:f2} shows the cumulative distribution of the number of dAGNs versus
the projected separation $r_{\rm p}$ of its two components for both observed
dAGNs and simulated dAGNs.  As seen from Figure~\ref{fig:f2}(a), the
simulated distribution matches well the observational one if $D_{{\rm c},i}$
is set to $\sim 1.25r_{{\rm h},j}$.\footnote{
We caution here that the total
number of dAGNs in \citet{Shen10} sample is only 5, and the separation
distribution of these dAGNs shown in Figure~\ref{fig:f2} suffers from small
number statistics. Future observations which promise to find more dAGNs would 
provide firmer constraints on $D_{\rm c}$ and would yield better statistics.
} 
After adding the dAGN candidate found by
\citet{Comerford09b} and the two dAGNs---major merger systems recently found by
\citet{Rosario11} to the dAGN sample, the dAGN $r_{\rm p}$
distribution appears to be still consistent with $D_{{\rm c},i}\sim (1-1.5)r_{{\rm h},j}$.
We choose $K=1.25$ as the reference value in this paper.  A much larger or
smaller $D_{\rm c}$ (e.g, $K=2$ or $0.5$) appears not to be consistent with the 
observationally-deduced distribution of projected
separations (see also the discussion on tidal interactions below). According to
equation (\ref{eq:tmrg}), the uncertainties
in the settings of $D_{\rm c}$ may lead to an error of about $30\%$
in the simulated LFs of the bright (or relatively faint) components of dAGNs.
Figure~\ref{fig:f2}(b) shows the evolution of the dAGN separation
distribution. The dAGNs at higher redshifts appear to have smaller separations,
which is mainly due to the size evolution of galaxies set in the model.  

\begin{figure}
%\epsscale{1.0}
\epsscale{0.8}
\plotone{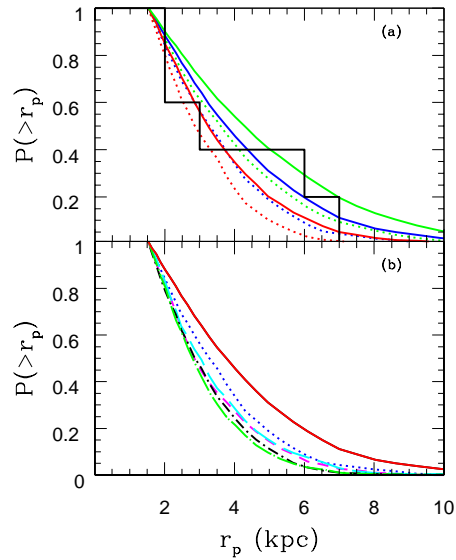}
\caption{The cumulative distribution of projected separation of dAGNs. The
histogram shows the observational results obtained by \citet{Shen10}. The
simulated dAGN sample is selected by the luminosity of the bright component with 
$L_{\rm [OIII]}>10^{7.5}\lsun$, as is set for the lowest luminosity threshold
of the observed dAGNs. The simulated separation distribution is normalized
to $r_{\rm p}=1.5\kpc$, as is done for the observational distribution \citep{Shen10}.  Panel (a)
shows the results of our simulations at redshifts $z=0.1$ (solid lines) and $0.3$
(dotted lines), respectively. The separation threshold for triggering nuclear
activities, i.e.,  $D_{{\rm c},i}=K r_{{\rm h},j}$ ($i=1,2$ and $i\ne j$ for
the two merging components)
is set to $K=1$ (red lines), $K=1.25$ (blue lines) and $K=1.5$ (green
lines). Panel (b) shows the evolution of the distribution of the projected separations
for the case of $K=1.25$. The red solid, blue dotted,
magenta dashed, cyan long-dashed, black dot-short-dashed, and green 
dot-long-dashed lines represent the results of our simulations at redshifts $z=0.1$, $0.3$,
$0.5$, $0.7$, $0.9$, and $1.1$, respectively. }
\label{fig:f2}
\end{figure}

\begin{figure}
%\epsscale{1.0}
\epsscale{0.8}
\plotone{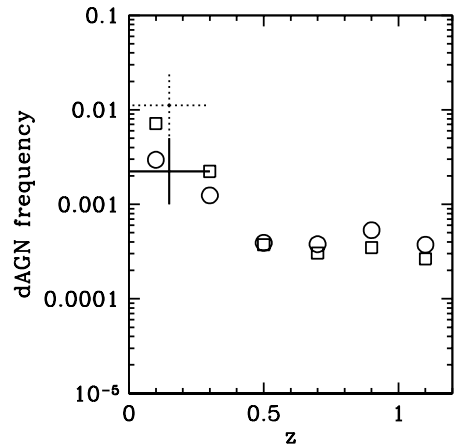}
%\centerline{\includegraphics[width=6.5in]{f3.eps}}
\caption{The expected dAGN frequency at different redshifts, i.e., the ratio of the number density of
simulated dAGNs to the number density of AGNs derived from the observational AGN LF.
Open circles and squares
represent the cases for the bright components of dAGNs and AGNs with
luminosities $L_{\rm [OIII]}\ge 10^{6.5}\lsun$ and $10^{7.5}\lsun$,
respectively. The dAGNs are selected through the double-peaked narrow lines in
our Monte Carlo simulations described in Section~\ref{sec:results}. In
this figure, the separation threshold for triggering nuclear activities is
set to have $K=1.25$. The typical error on the estimated dAGN frequency is $0.5-0.6$~dex (see
discussions in Section~\ref{sec:results}).
The total
[OIII] LFs of AGNs at different redshift are adopted from \citet{Bongiorno10}.
The dotted and solid crosses are the current observational constraints obtained with and without completeness correction, respectively. 
See details in Section~\ref{sec:results}.  }
\label{fig:f3}
\end{figure}

\item The accretion rates of dAGNs are set to have $\lambda=0.25$ times the
Eddington rate, similar to those in observed bright QSOs (see
Section~\ref{subsec:lumevol}). If $\lambda$ is set to be $0.1$, the LFs of the
simulated dAGNs move downward by a factor of $2-3$ at $L_{\rm [OIII]}
\sim 10^7-10^8\lsun$; and if $\lambda=1$, the LFs of the simulated dAGNs increase
by more than a factor of 10 at $L_{\rm [OIII]}\ga 10^{8.5}\lsun$ but 
stays the same and then decreases slightly as $L_{\rm [OIII]}$ decreases
to less than $10^{7.5}\lsun$.

\item The relative velocity $v_{\rm rel}$ of the two components of any merging
pair is set to be twice the circular velocity of their common dark
matter halo. This may be an overestimate of $v_{\rm rel}$ as the primary
component may move at a lower speed than the circular velocity. However, it is
hard to address the uncertainty in the dAGN LF caused by this
assumption. In order to get a sense of this uncertainty, here we simply assume
a case of $v_{\rm rel}\sim 1.5 v_{\rm circ}$ or $v_{\rm circ}$ and find
that it could lead to a decrease of the dAGN LFs by about a factor of 1.4 or 2.

\end{itemize}

Combining all these uncertainties together, the total systematic uncertainty
in the LFs of the bright components of dAGNs is likely to be a factor of $3-4$
over the luminosity range $10^6-10^{8.5}\lsun$. 

Figure~\ref{fig:f3} shows the evolution of the dAGN frequency, i.e., the ratio
of the number density of simulated dAGNs selected through the double-peaked narrow line features to the number density of AGNs derived
from the observational total AGN LF \citep{Bongiorno10}.
Here we do not distinguish between type 1 and type 2 AGNs, but include both.
This ratio
is equal to the ratio of the cumulative LF of dAGNs to that of AGNs, and
depends weakly on the lower limit of the AGN luminosities.  These
estimates could be off by a factor of $3-4$ as previously discussed. At redshift $z\sim
0.1-0.3$, the estimated dAGN frequency is $\sim 0.1\%-1\%$. (This value could be higher than the estimates from observed dAGNs
selected through the double-peaked narrow lines, considering that the signal-to-noise of some dAGN spectra
are so low that the dAGNs are not detected observationally.)
There are quite a number of 
dAGNs which cannot be detected by using double-peaked narrow lines, due to the projection effect and the velocity cutoff. Including the undetected dAGNs, our model suggests that
the real frequency
of dAGNs is $\sim 0.2\%-2\%$, larger than that of those dAGNs selected through the double-peaked
narrow line features by a factor of $\sim 2$. 
These values are fully consistent with the
observational estimates obtained by \citet{Shen10}, i.e., $\sim 0.1\%-0.5\%$ without 
completeness correction or $\sim 0.5\%-2.5\%$ after the correction of in-completeness due to the projection effect, velocity cutoff and low signal-to-noise spectra of some dAGNs (see also
\citealt{Rosario11}, in which the dAGN frequency is $\sim 0.4\%$ without the completeness correction).

The reasons that a lower frequency of dAGNs is obtained from our model
(compared to the simple estimates; e.g., $15\%$ by \citealt{Rosario11}) are
three folds: (1) dAGNs can be produced only if both of the two progenitor
galaxies are gas rich so that sufficient accretion materials could be provided;
(2) MBHs in those gas-rich progenitor galaxies may be initially substantially
smaller than their final masses, and thus the AGN phenomena triggered before
MBH mergers may be substantially less luminous compared with those triggered
after MBH mergers; and (3) in a significant fraction of major mergers, the
masses of the MBHs hosted in the two progenitor galaxies may differ by
a large factor and thus the induced dAGNs may have too large luminosity
difference to be included. 

As shown in Figure~\ref{fig:f3},
the modeled frequency of kpc-scale dAGNs that can be detected through the double-peaked narrow line features
declines to even lower values on the order of several
ten-thousandths, e.g., $0.02\%-0.06\%$ at redshift $z\sim 0.5-1.2$. And the frequency
of all kpc-scale dAGNs declines to $0.04\%-0.1\%$ at redshift $z\sim 0.5-1.2$. 
This decline is mainly due to the redshift evolution of the galaxy morphology
distribution \citep[e.g.,][]{Zucca06}. Correspondingly, the major mergers at such redshifts 
are dominated by very late-type galaxies (e.g., Sc-Sd and Irr galaxies) and the 
fraction of mergers that host large initial MBHs becomes substantially smaller at 
high redshifts.  Future observations which would specifically search 
for high redshift dAGNs could test this prediction.

In addition, we note here that the estimates of the dAGN LF and the separation 
distribution may be affected by some simplified assumptions made above. Below
we discuss how our results would be affected by some other mechanisms, which may
trigger nuclear activities, but is ignored in the model above.

First, we have explicitly assumed that significant nuclear activities
can only be triggered in the gas-rich component of a major  merger. If we relax
this assumption and allow significant nuclear activities to be triggered
in the gas-poor component of a major merger, then the estimated dAGN frequency would rise
to about ten percent or higher at luminosity $L_{\rm [OIII]}\ga 10^{8.5}L\sun$,
but is not affected much at lower luminosities. Under this assumption, the dAGNs at the
high-luminosity end are primarily formed by major mergers of two red progenitor
galaxies as the initial MBHs in red galaxies are substantially big compared to
those in blue galaxies with similar total stellar masses. Mixed mergers of a red progenitor
galaxy with a blue progenitor galaxy contribute little to the population of dAGNs
with comparable luminosities, as the masses of the initial MBHs in these two progenitor
galaxies are less likely to be comparable.

Second, a dAGN can also be produced at larger separations if nuclear activities could 
be triggered by tidal interactions when the two progenitor galaxies of a major merger
are still far away from each other. Under this assumption, the model described
in Section~\ref{sec:model} can still be used but a large value of $K$ should be chosen
(e.g., $K\sim 5-10$). By assuming $K\sim 5-10$ and a constant Eddington ratio $\lambda=0.25$
till the completion of the mergers, our estimate of the  frequency of
kpc-scale dAGNs increases to $\sim 10\%$. This increase
is clearly because most of the MBHs
in the progenitor galaxies could now grow up significantly before the separation of the two
components of a dAGN decays to the kpc scale.  
However, increasing the value of the $K$ parameter to $\sim 10$ also implied 
that the predicted number of AGN pairs on scale of $10-100\kpc$ from our model is comparable to the total
number of AGNs at luminosity $L_{\rm [OIII]}\ga 10^{7}L\sun$, which is
not consistent with current observations. One possible way to solve this inconsistency
would be to impose that tidal interactions induce only 
low-level nuclear activities (e.g., with $\lambda \sim 0.01-0.001$) when the two 
progenitor galaxies are still far away from each other and the accretion rate is
significantly enhanced only when the two progenitor galaxies become close enough
to each other. In this case, a large number of paired AGNs on $10\kpc-100\kpc$ (with 
a frequency of several percent to ten percent) can be produced but the resulted
frequency of dAGNs on kpc-scale is still the same as the current observations suggest.

Third, successive mergers, in principle,  can include mergers
of two progenitor galaxies whose nuclei have already been activated,
which would also yield a dAGN. This scenario has not been discussed
here in details mainly for simplicity. However, according to the major merger rates of galaxies presented in 
Section~\ref{sec:model}, the rate of successive major mergers, 
within the lifetime of detectable nuclear activities (e.g., $<1\Gyr$),
is extremely low, especially if successive merger events were assumed
to be independent of each other.

Future observations which would determine the distribution of dAGNs as a function of
luminosity and separation may shed light on our understanding of
the triggering mechanisms,  major mergers, tidal interactions
and the mergers of galaxies with already-activated nuclei.

\section{Conclusions}\label{sec:conclusions}

In this paper, we have constructed a phenomenological model to estimate the number
density of dAGNs and its evolution according to the observationally-estimated
merger rates of galaxies and various scaling relations on the properties of
galaxies and massive black holes. We have demonstrated that the observed frequency
and separation distribution of dAGNs are compatible with the hypothesis that
major mergers of galaxies lead to AGN/QSO activities and are consistent with
most estimates of galaxy major merger rates. We have also predicted that the 
frequency of those dual AGNs with double-peaked narrow line features becomes even smaller (i.e., $0.02\%-0.06\%$) 
at higher redshifts $z\sim0.5-1.2$ and the frequency of all kpc-scale dAGNs is $\sim 0.04\%-0.1\%$ at $z\sim 0.5-1.2$, which can be tested by future observations. 
Future observations shall provide us with much better 
statistics on the distribution of kpc-scale dAGNs (and pairs
of AGNs on even larger scale) as a function of
luminosity and separation and hence will tighten the constraints 
on triggering mechanisms of nuclear activities.

\acknowledgements
We thank Yue Shen for helpful comments.
This work was supported in part by the LIA: Origins International Associated
Laboratory between China and France in astronomy, by the National Natural
Science Foundation of China under No.\ 10973001, 10973017, 11033001,  by
the Bairen program from the National Astronomical Observatories, Chinese
Academy of Sciences, and by French ANR OTARIE.

\end{document}